\documentstyle[11pt,aaspp4]{article}

\begin{document}
\title{The Tip of the Red Giant Branch Distance to the Large Magellanic Cloud}

\medskip
\begin{center}

Shoko Sakai$^{1}$,
Dennis Zaritsky$^{2}$,
Robert C. Kennicutt, Jr.$^{2}$
\end{center}

\altaffiltext{1}{Kitt Peak National Observatory P.O. Box 26732, Tucson, AZ 85726}

\altaffiltext{2}{Steward Observatory, University of Arizona Tucson, AZ}

\begin{center}
{\it Accepted for Publication in the Astronomical Journal, March 2000}
\end{center}

\def\Deg{\hbox{${}^\circ$\llap{.}}}
\def\Min{\hbox{${}^{\prime}$\llap{.}}}
\def\Sec{\hbox{${}^{\prime\prime}$\llap{.}}}

\begin{abstract}

We present the I--band luminosity function of the red giant branch stars
in the Large Magellanic Cloud (LMC) using the data from the Magellanic
Clouds Photometric Survey (Zaritsky, Harris \& Thompson, 1997).
Selecting stars in uncrowded, low--extinction regions, 
a discontinuity in the luminosity function is observed at $I_0 = 14.54$ mag.
Identifying this feature with the tip of the red giant branch (TRGB),
and adopting an absolute TRGB magnitude of $-4.05 \pm 0.04$ mag based on the
calibration of Lee, Freedman \& Madore (1993),
we obtain a distance modulus of $18.59 \pm 0.09$ {\small {(random)}} $\pm 0.16$
{\small {(systematic)}} mag.
If the theoretical TRGB calibration provided by Cassisi \& Salaris (1997) is
adopted instead, the derived distance would be 4\% greater.
The LMC distance modulus reported here, $18.59 \pm 0.09$,
is larger by $0.09$ mag (1$\sigma$)
than the value that is most commonly used in the extragalactic distance scale
calibrated by the period--luminosity relation of the Cepheid variable stars.
Our TRGB distance modulus agrees with several RR Lyrae
distances to the LMC based on HIPPARCOS parallaxes.  
Finally, we note that using the same MCPS data, we obtain
a distance modulus of $18.29 \pm 0.03$ mag using the
red clump method, which is shorter by 0.3 mag compared
to the TRGB estimate.

\end{abstract}

\it Subject headings: \rm Magellanic Clouds   --
galaxies: irregular galaxies -- galaxies: distances
\bigskip
\bigskip

\section{Introduction}

The distance to the Large Magellanic Cloud (LMC) continues to serve as the
cornerstone in the extragalactic distance scale.
Distances to external galaxies measured using the Cepheid variable star
period--luminosity (PL) relation are usually determined with respect to 
the distance of the LMC.
Recently, Mould et al. (2000) reported the final results from the
Hubble Space Telescope (HST) Key Project on the Extragalactic Distance
Scale: H$_0$ = 71 $\pm$ 6 km s$^{-1}$ Mpc$^{-1}$.
One of the largest contributors to the systematic error is the uncertainty
in the LMC distance modulus, assumed to be $18.50 \pm 0.13$ mag (Mould et al. 2000).
The uncertainty in this value was derived by examining the distribution of
various LMC distance estimates published in literature.
As will become clear below, it is not surprising that the distance
to LMC is not determined more precisely.

Distance estimates for the LMC span from 18.1 to 18.7 mag.
For details of this field, readers are referred to discussions in
Westerlund (1997), Jha et al. (1999), and Mould et al. (2000).
Here, we restate some of the latest results, many of which
were derived using HIPPARCOS parallax measurements.
Feast \& Catchpole (1997)  reported a dereddened distance modulus of
$(m-M)_0^{\tiny LMC} = 18.70 \pm 0.10$ mag,
using the HIPPARCOS distances to Galactic Cepheid variables, which were
then compared with the LMC Cepheids from Caldwell \& Laney (1991).
However, three independent re--analyses suggest that the
distance modulus derived by Feast \& Catchpole can be lowered by as 
much as $\sim 0.35$ mag
(Madore \& Freedman 1998, Oudmaijer, Groenewegen \& Schrijver 1998,
Luri et al. 1998).  Likewise HIPPARCOS parallaxes to Galactic
subdwarfs have been used to redetermine the RR Lyrae distance to the
LMC, yielding $(m - M)^{LMC}_0$ = 18.65 mag (Reid 1997), 18.60 $\pm$ 0.07 mag
(Gratton et al. 1997), 18.54 $\pm$ 0.04 mag (Carretta et al. 1999),
and 18.61 $\pm$ 0.28 mag (Groenewegen \& Salaris 1999).  
A HIPPARCOS based calibration of Mira variables
by van Leeuwen et al. (1997) yields an LMC distance modulus of 
18.54 $\pm$ 0.2 mag.  These results imply LMC distances that are 2--10\%\
higher than the conventional value of 50 kpc [$(m - M)_0$ = 18.50].
Conversely, evidence for a shorter distance comes from a relatively new
method that utilizes the luminosity of red clump stars (Udalski 1998,
Stanek, Zaritsky, \& Harris 1998).  The derived distance modulus is
sensitive to corrections for age and metallicity differences between
the Galactic calibrating stars and the LMC, and values ranging from
$(m - M)_0$ = 18.08 to 18.36 have been reported (Stanek et al. 1998,
Udalski 1998, Cole 1998, Girardi et al. 1998).  Recently Zaritsky (1999)
reanalyzed the reddening distribution in the LMC, taking into account
differences in reddening as a function of stellar color, and he concludes
that the published red clump distance moduli need to be revised upwards by
$\sim$0.2 mag (to $\simeq$18.3--18.55 mag) to correct for an overestimate 
in the previously adopted reddening values.  
Romaniello et al. (1999) also obtained a low extinction to the red clump
stars.  They further applied a theoretical metallicity correction to obtain 
a larger distance modulus of 18.59 mag.
All of these values can
be compared to the geometric distance measurement to the LMC based on
the SN 1987A expanding ring, which yields $\mu_0^{LMC}$ = 18.44 -- 18.58
mag (Gould \& Uza 1998, Panagia et al. 1997).
Unfortunately, convergence on a precise distance still eludes us.

Zaritsky, Harris \& Thompson (1997: hereafter ZHT) have been undertaking the Magellanic
Clouds Photometric Survey (MCPS) using the Las Campanas 1m telescope.
A roughly 4$^{\circ}$ by 2.7$^{\circ}$ region of the LMC has been analyzed to date using
the U-, B-, V- ,and I-band filters.
This database presents us with an opportunity to re-measure the
distance to the LMC -- this time, using the tip of the red giant branch (TRGB)
method, which uses the I--band luminosity function of the red giant
branch stars.
These stars evolve upward on the red giant branch, but undergo a drastic
physical change at the onset of the core helium flash.
The method is currently calibrated using the distances to Galactic
globular clusters (Da Costa \& Armandroff 1990, Lee, Freedman \& Madore 1993).
The TRGB method is powerful because as a bright Population~II distance indicator,
it is applicable to any morphological type of galaxy,
and red giants are found in abundant numbers, yielding good number statistics.
In addition, one can select the RGB stars strategically by avoiding
regions with dust and young stars, 
thereby minimizing uncertainties due to crowding and reddening corrections.

An earlier application of the TRGB method to determine the LMC distance
was reported by Reid, Mould \& Thompson (1987), who examined
the stellar populations of Shapley Constellation~III in the LMC.
They detected the TRGB at $I = 14.60 \pm 0.05$ mag.
By adopting a bolometric correction of 0.39 mag (corresponding to M stars),
extinction of $A_I = 0.07$ mag, 
and an absolute bolometric magnitude of the TRGB of $\sim -3.5 \pm 0.1$ 
implied from Frogel, Cohen \& Persson (1983), Reid et al. (1987) reported
$(m-M)_0^{\tiny {LMC}} = 18.42 \pm 0.15$ mag.  
More recently, Romaniello et al. (1999) used the HST/WFPC2 multiband observations
of a field around SN1987A to estimate the magnitude of the TRGB.
Using a theoretical calibration of Salaris \& Cassisi (1998), they obtained
the distance modulus of $(m-M)_0^{LMC} = 18.69 \pm 0.25$ mag.

We re-determine the distance to the LMC using the TRGB
method, using the significantly larger photometric database of ZHT.
Because this method is based on a distance scale that is completely 
independent of the Pop~I Cepheid distance indicators, it can provide
an alternative Pop~II check on the distance to the LMC.
The data used for the analysis presented in this paper are described
in \S2.  We then discuss the derivation of the TRGB distance
(\S3), followed by a discussion of the errors (\S4), and a summary of
our conclusions (\S5).

\section{The Data}

The data used for this study come from the ongoing MCPS described initially by ZHT.
The survey is being conducted using the Las Campanas
Swope (1m) telescope with the Great Circle Camera (Zaritsky, Shectman, \& Bredthauer 
1996) and a 2K by 2K CCD with a pixel scale of 0.7" pixels.
The effective exposure time per filter is about 4 min. Details of the data reduction are
described by ZHT and Zaritsky (1999). The region of the LMC 
being studied here is roughly 4$^{\circ}$ by 2.7$^{\circ}$ centered at $\alpha
=5^h20^m$ and $\delta=-66^\circ 48^\prime$. Roughly 4 million
stars are photometered in both the B- and V-band images (about half as many
in $U$ and $I$).

We recover the reddening along the line of sight
to individual stars as described in detail by Zaritsky (1999). We fit
stellar spectra plus extinction to stars in our survey with $UBVI$ photometry.
We recover both an estimate of the effective temperature
of the star and $A_V$, for the assumed standard extinction curve. We find that 
the technique is reliable only for stars with effective temperatures
5500 K$< T_{eff} <$ 6500 K and $T_{eff} > 12000$ K. For stars with other effective temperatures,
the recovered $T_{eff}$ and $A_V$ are sufficiently degenerate that the results of the
fitting algorithm are suspect. The lower of the two temperature 
ranges includes the LMC red clump stars and some of the giant branch. Because
the distribution of extinction values is different for the hotter 
and cooler stars, it is critical to measure the reddening
to a population that is as similar as possible as the population being studied.
Therefore, to derive reddenings for the TRGB stars we choose to use those derived
from the 39,613 stars that are in the cooler temperature range.

To correct the TRGB star photometry, we convert our extinction
measurements for individual stars to a map of extinction across the 
region. Because the extinction to any individual star is significantly 
uncertain ($\sigma \lesssim 0.2$ mag) we average local values and interpolate to 
produce our extinction map. Again, details and tests of the procedure
are described by Zaritsky (1999). The map has a variable
spatial resolution that depends on the local surface density of extinction
measurements, with the highest spatial resolution of 84\arcsec. 
The reddening corresponding to each resolution element is determined from at least 3 stars, 
so typical formal uncertainties per resolution element are $\sigma_{A_V} <$ 0.1 mag
(thus $\sigma_{A_I} < 0.05$ mag).

\section{Tip of the Red Giant Branch Distance to the LMC}

\subsection{TRGB Calibration}

The TRGB marks the core
helium flash of first--ascent red giant branch stars. 
In the I--band, the TRGB magnitude is 
insensitive to both age and metallicity (Iben \& Renzini 1983).
In composite stellar populations, it 
is observed as a sharp discontinuity in the I--band luminosity
function.  Currently, the TRGB calibration is based on RR Lyrae
distances to six Galactic globular clusters spanning a range of metallicities
$-2.1 \leq$ [Fe/H] $\leq -0.7$, estimated using the metallicity-$M_V$
correlation by Lee, Demarque \& Zinn (1990) (Da Costa \& Armandroff 1990: hereafter DA90,
Lee, Freedman \& Madore 1993:hereafter LFM93).  

Using the I--band TRGB magnitude ($I_{TRGB}$), the distance
modulus is estimated via a relation:
\begin{equation}
(m-M)_I = I_{\tiny {TRGB}} + BC_I - M_{\tiny {bol,TRGB}}.
\end{equation}
Both $BC_I$ and $M_{\tiny {bol,TRGB}}$ terms are functions of the metallicity (LFM93) :
\begin{equation}
M_{\tiny {bol,TRGB}} = -0.19 [Fe/H] - 3.81,
\end{equation}
\begin{equation}
BC_I = 0.881 - 0.243(V-I)_{\tiny {TRGB}}.
\end{equation}
The bolometric correction was derived by DA90 by comparing their optical
photometry of globular clusters
with the IR photometry from Frogel, Persson \& Cohen (1983).
The metallicity, [Fe/H], is determined from the 
de--reddened $(V-I)$ colors of the RGB stars:
\begin{equation}
[Fe/H] = -12.64 + 12.6 (V-I)_{\tiny {-3.5}} - 3.3 (V-I)^2_{\tiny {-3.5}},
\end{equation} 
where $(V-I)_{\tiny {-3.5}}$ is measured at the absolute I magnitude of $-3.5$

The calibration presented by equations 1--4 
is semi--empirical, based on an RR Lyrae
distance scale using the theoretical models of horizontal branch stars.
Cassisi \& Salaris (1997) recently presented a purely theoretical calibration
of the TRGB method, by examining their stellar evolution models (Salaris \&
Cassisi 1996). 
The main result of that work is that the theoretical calibration is $\sim$0.1
mag brighter than the empirical calibration.  
They attribute this difference to a poor sampling of the RGB
stars in the Galactic globular clusters used for the DA90 calibration,
and the small sample of stars observed by Frogel, Persson \& Cohen
(1983) whose photometric data were used to estimate the bolometric corrections.
In \S4.2, we discuss how the uncertainties in 
the calibrations affect out LMC distance estimate.

\subsection{TRGB Magnitude}

In Figure~1, we show the reddening corrected
$I-(V-I)$ and $I-(B-I)$ color--magnitude diagrams
for stars in the upper RGB region for the entire LMC region surveyed.  
In order to make a reliable TRGB detection, we select RGB stars
using the following criteria:

(1) Stars in the shaded region in Figure~1 are excluded, as their magnitudes
and colors are obviously inconsistent with RGB stars.  Also, those with
$I \geq 16.0$ were not included to minimize the computational time.

(2) Using the extinction values derived from stars of effective temperatures
$T_{eff} > 12,000$ K as described in Section~2, 
those regions with $A_V \geq 0.2$ are excluded,
to avoid dusty regions.

(3) Any stars that lie in a crowded region are excluded.  This was
done by excluding those stars lying in a region for which the density map value,
$\sigma$, is greater than 1.0 star/pixel.
The stellar density map was constructed by counting stars with $V < 21$ in 21\arcsec\ pixels. 

\noindent
We are left with a sample of 2072 stars, consisting mostly of
RGB stars and some intermediate--age asymptotic giant branch stars
(those brighter than $I \sim 14.5$).

In Figure~2 we show the corresponding I--band luminosity function
from $M_I = 13$ to 16 mag {\it (top)}, 
and the output of the edge--detection filtering of the I--band luminosity
function.  The details of the edge--detection filter were described by
Sakai, Madore \& Freedman (1997).  
The technique is designed to pick out the luminosity at which the slope reaches
the maximum.
Briefly, the filter is a modified 
Sobel kernel, $[-1, 0, +1]$, and it is applied to a 
luminosity function that has been smoothed using a Gaussian of dispersion equal
to the photometric error.
The position of the TRGB is indicated by the highest peak in the lower
panel of Figure~2.  We detect the TRGB at $I_0 = 14.54 \pm 0.04$ mag.
The error quoted here, 0.04 mag, refers to the FWHM
of the peak profile in the filter output function.
We note that our TRGB magnitude is 0.06 mag brighter than
the value of $14.60 \pm 0.05$ derived by Reid et al. (1987).
Although the discrepancy is slight ($\sim$1.5$\sigma$),
it may reflect the sampling bias.
Romaniello et al. (1999) detected the TRGB using the HST/WFPC2 observations
at $I_0$ = 14.50 $\pm$ 0.25 mag,
which agrees with our value well within 1$\sigma$.
However, they obtain a larger distance modulus of 18.69 $\pm$ 0.25 mag,
mainly due to their use of a theoretical (Salaris \& Cassisi 1998) rather
than the empirical calibration of Lee et al. (1993).

From the RGB sample used to determine this TRGB magnitude, 
we find $\overline{(V-I)}_{\tiny {TRGB}} = 1.7 \pm 0.1$ mag
and $\overline{(V-I)}_{\tiny {-3.5}} = 1.5 \pm 0.1$ mag.
Substituting these colors into the calibration formulae,
we obtain $M_{\tiny {I,TRGB}} = -4.05 \pm 0.06$ mag.
Thus, the distance modulus to the LMC determined by the TRGB method
is $(m-M)^{LMC}_0 = 18.59 \pm 0.07$ mag, corresponding to the linear
distance of $52 \pm 2$ Kpc.
The error quoted here only includes random terms:
0.04 mag uncertainty in the edge--detection method, and
0.06 mag from the color spread in the RGB population.
We discuss the systematic errors in the distance in \S4.

\section{Uncertainties in the TRGB Distance to the LMC}

In Table~1, the uncertainties in the LMC distance modulus are summarized.
They include those due to the photometric and reddening estimates,
the TRGB calibration zero point, and the effects of the line--of--sight
depth of the LMC itself.
We review in the following subsections some of the uncertainties in detail.

\subsection{Photometry and Reddening}

Internal photometric errors are calculated by DAOPHOT~II (Stetson 1987). 
From comparison
of results from overlapping images, we conclude that the internal uncertainties 
at worst underestimate the true uncertainties in the instrumental magnitudes by 
a factor of 1.5 (ZHT). We have multiplied the DAOPHOT uncertainties by this 
factor to be conservative. Because of the large number of stars ($\sim$1000)
used in determining the TRGB position, the internal photometric errors 
have an insignificant effect on the fitted distance modulus.
The uncertainty in the zero point of the photometric scale
appears small ($\leq 0.05$ mag) as well from some 
comparisons to previous data; as will be shown later in Section~6,
there is good  agreement between the red clump magnitude obtained
here with independent measurements and also between our TRGB magnitude with
that from Reid et al. (1987).

The median $A_V$ in a single resolution element is determined to $\pm$0.1 mag. Small
scale reddening structures are not resolved and so the uncertainty in the extinction 
of any single star can be much larger. However, all of our measurements are based on 
large samples, where mean extinction
should be well determined from the map. We have used the uncertainty of 0.1 mag per 
resolution element and calculated the number of resolution elements included in the 
selected regions to determine the error in the mean extinction of the region to be $\sim
\pm$ 0.03 mag.

\subsection{Uncertainties in the TRGB Calibration}

The TRGB calibration used in
this paper is based on the  distances to six Galactic
globular clusters which were determined by the metallicity--$M_V$
relation for RR Lyrae variable stars given by Lee, Demarque \& Zinn
(1990) for $Y_{\tiny {MS}} = 0.23$, expressed as $M_V(RR) = 0.17 [Fe/H] + 0.82$.
Unfortunately, both the zero point and the degree of metallicity dependence
is uncertain.  
In particular, the range of zero point estimates by different groups 
is as much as 0.3 mag.
Chaboyer (1999) has published an excellent review on the RR Lyrae distance
scale, in which he calibrated the method
using several different methods that included statistical parallax fitting,
theoretical horizontal branch models, and main sequence fitting using
the HIPPARCOS database.
For the LMC, where [Fe/H] = $-1.18$, adopting the five different
calibrations given by Chaboyer (1999), we obtain RR Lyrae
magnitudes, M$_V$(RR), the range from 0.55 to 0.89 mag with the average value
of 0.68 $\pm$ 0.11 mag.
The RR Lyrae calibration of Lee et al. (1990), which we adopted in the TRGB calibration 
would yield, for [Fe/H] = $-1.18$, M$_V$(RR) = 0.62, which is in agreement with the
mean of these more recent calibrations.  We therefore adopt a systematic
error in the RR Lyrae distance scale of 0.11 mag.
 
A purely theoretical calibration of the TRGB method was presented recently
by Cassisi \& Salaris (1997:hereafter CS97) who used the evolutionary models
of stars for a combination of various masses and metallicities for
$Y_{\tiny {MS}} = 0.23$ (Salaris \& Cassisi 1996). 
CS97 find that the empirical zero point of the TRGB calibration is fainter by $\sim$0.1 mag
than that of the theoretical calibration, mainly due to (1) sampling errors:
not enough stars
populate the TRGB region in the Galactic globular clusters used in the
empirical calibration, such that the probability of actually observing
the brightest TRGB stars is very small;  and (2) statistical uncertainties introduced
by the small number of stars  used by
Frogel, Persson \& Cohen (1983) to derive the bolometric correction.
The importance of sampling the TRGB was confirmed empirically by
Sakai \& Madore (1999), who reported that when only one in five stars of the true
stellar population was used in estimating the position of the TRGB, 
its magnitude became fainter  by $\sim 0.06$ mag.
The CS97 calibration is given as follows:
\begin{equation}
M_{\tiny {I,TRGB}} = -3.953 + 0.437 [Fe/H] + 0.147 [Fe/H]^2,
\end{equation}
\begin{equation}
[Fe/H] = -39.270 + 64.687 (V-I)_{-3.5} - 36.351 (V-I)^2_{-3.5} + 6.838 (V-I)^3_{-3.5}.
\end{equation}
If we were to adopt this calibration instead of that of
Lee et al., we obtain $M_{\tiny {I,TRGB}} = -4.13$ mag, and thus
the distance modulus of $(m-M)^{LMC}_0 = 18.67$ mag,
$\sim$4\% further than the value derived using Lee et al. calibration.
Here, we adopt an uncertainty of 0.1 mag as a systematic error in the
calibration due to small globular cluster samples.

Another possible uncertainty in the TRGB calibration originates from
the fact that the zero--point calibration of LFM93 did not use the 
edge--detection filtering. Because the method used to determine the
TRGB magnitudes for Galactic globular clusters is different from
that used to measure the extragalactic TRGB magnitudes, there could 
be a systematic difference in the results.
Unfortunately, because the globular cluster data are much less densely populated
than most other extragalactic data,
it is difficult to securely measure the TRGB magnitude
using the filtering method.  
Nevertheless, we apply the edge filter
to the globular cluster data to examine whether we observe any
large systematic errors.  By combining the data from all six globular clusters,
we obtain a TRGB magnitude of $-4.05 \pm 0.03$ mag.
This result agrees precisely with the zero point we have adopted,
suggesting that there is no major
systematic error from inconsistent TRGB measuring techniques between
the calibrators and the galaxies.

We have adopted an error of 0.06 mag due to the color spread in the LMC
RGB population (see Table~1). However, we feel that this is a very
conservative estimate.
Using the LMC data, we find that the TRGB magnitude
is insensitive to the $(V-I)$ color range sampled.
Dividing the RGB sample into red and blue subgroups, 
the TRGB magnitude is the same for both of them.
This result confirms that adopting one zero point for the entire TRGB star 
population that spans a broad range in color should not make the TRGB
distance modulus estimate any more uncertain.

\subsection{Effects of Crowding and Extinction on the TRGB Distance Determination}

When applying the edge--detection filter to the I--band luminosity function to 
determine the TRGB magnitude, one usually avoids the central region of the galaxy
where the extinction is large and the crowding becomes a severe problem.  
Because the TRGB method requires an independent estimate of the internal extinction
of the galaxy, the degree of extinction, and the uncertainty in the adopted
value, must be minimized.
For our analysis of the LMC, we are fortunate to have
excellent extinction maps, determined from two sets of stars:
those with temperatures higher than 12,000 K;
and those with effective temperatures between 5,500 and 6,500 K.
As discussed earlier, the extinction map inferred from the hot stars 
was used to exclude the high--extinction regions.
Here, we examine whether the TRGB magnitude is sensitive to this particular
selection, by showing in Figure~3 the I--band luminosity function and
corresponding edge--detection filter outputs for four different
extinction cutoff values.  
The number of stars used in each subsample is indicated in brackets.
Most of the stars in high--extinction regions also lie
in more crowded regions.  
For the bottom case (highest extinction cutoff), even though
the TRGB position is visible as the highest peak in the
filter output function, it is barely visible in the luminosity function.

Similarly, we use the density map to examine how crowding affects
the determination of the TRGB magnitude.
In Figure~4 the luminosity functions and filter output functions
are shown for four different surface density ranges
(density ranges increasing from top to bottom).  
We apply the standard extinction cutoff, $A_V \leq 0.3$ mag, 
to all of the samples.
Figure~4 illustrates the rapid degradation of the TRGB detection at high
stellar density, which justifies our exclusion of regions with density $>1.0$.

\subsection{The Line--of--Sight Depth and the Tilt of the Large Magellanic Cloud}

If the bulk of the LMC population subtends
a large depth along the line--of--sight due either to an intrinsic
thickness or high inclination, the fitting of the TRGB
might tend to preferentially sample the nearest (and brightest)
giants, and thus lead to a systematic underestimate of the LMC distance.
We explore this effect using a simple simulation.  We also examine
whether these simulations will help in estimating the actual line--of--sight
depth of the LMC.

We begin with a luminosity function that is roughly defined as
$N(M_I<-4) = 0$, and $N(M_I \geq -4) \propto M^{0.6}$.
About 7500 stars were generated in the magnitude range $-4 \leq M_I \leq -3$.
Then random photometric noise following a Gaussian distribution with a FWHM of 0.03 mag
was added to each magnitude to simulate the observed photometric error
in the true LMC data.
This distribution simulates a sample with zero line--of--sight depth.
We then displaced each star along the line--of--sight randomly following a boxcar 
distribution,
and for each simulation, the TRGB magnitude and the dispersion of the profile
corresponding to the TRGB in the edge--detection filtering output were measured.
For boxcar back--to--front depths of 0.05, 0.10, 0.15, and 0.20 mag,
we obtained TRGB magnitudes of $-4.052 \pm 0.003$, $-4.033 \pm 0.012$,
$-4.016 \pm 0.021$, and $-4.007 \pm 0.026$ mag respectively.
Contrary to one's naive expectation, the TRGB magnitude becomes slightly fainter as
the line--of--sight depth becomes larger.
A similar trend is observed when a Gaussian line--of--sight distribution
is assumed instead of a boxcar.
The result of the broadening is translated into a smoothing of the luminosity
function.  
Because the luminosity function is steeper towards magnitudes fainter than the
TRGB than towards brighter magnitudes, the largest derivative,
which is what the filtering selects, is displaced to fainter
magnitudes.

From the same set of simulations, we calculate that the dispersion of the TRGB filter 
peak ``profile'' is $0.023 \pm 0.003$, $0.032 \pm 0.012$, $0.046 \pm 0.022$,
and $0.057 \pm 0.028$.  
Our measured profile width for the LMC is $\pm$0.04 mag.
This suggests that the LMC depth along the line
of sight lies in the range $0.10-0.20$ mag, which corresponds to 2.3--4.6 Kpc at a
distance of 50 Kpc.  
This range probably overestimates the actual depth because other sources of 
uncertainty, such as reddening errors, will contribute
to the width of the TRGB feature.
Xu, Crotts \& Kunkel (1995) studied the structure of interstellar
medium around the SN 1987A in the LMC and suggested that it extended
for up to $\sim$1Kpc. 
We can also put an upper limit to the LMC line--of--sight depth from
the magnitude distribution of red clump stars which is shown in Figure~5.
We observe a dispersion of $0.200 \pm 0.003$ mag, corresponding to 10\% in distance 
($\sim$5 Kpc).
This can be interpreted as a very conservative upper limit on the
LMC depth, and it compares well with the values obtained from the TRGB study.
Bessell, Freeman \& Wood (1986) found a much thinner disk
(scale height of only 0.3 Kpc), which is certainly consistent with our upper limits.
We note however that a variation in reddening could also affect and
dominate the width of the filter peak profile, leaving little room
for the thickness of the LMC itself.
For the purpose of this study, the relevant conclusion is not that we obtain
5 Kpc thickness, but rather that even when our simulations are run with such
a large upper limit, our determination of the TRGB magnitude is unaffected.

The tilt of the LMC with respect to the plane of the sky
would contribute an additional systematic uncertainty, in a very similar fashion 
as the observable line-of-sight
depth of the LMC disk that was discussed in the previous section.
Several papers have dealt with this issue previously, suggesting the tilt
angle of $\sim$25$^{\circ}$ up to $\sim$55$^{\circ}$ (de Vaucouleurs 1980,
Caldwell \& Coulson 1986, Laney \& Stobie 1986, Welch et al. 1987).
The MCPS survey region that was used to determine the TRGB distance
in this paper is located about 3$^{\circ}$ almost directly north from the center of the LMC,
which turns out to be a fortuitous situation since the line-of-nodes roughly bisects this
surveyed region;
an HI mapping of the LMC suggest that the line of nodes has the position angle
of $-12^{\circ}$ (Kim et al. 1998).
Thus, the TRGB distance measured in this paper should be close to
that of the center of the LMC, and
any systematic uncertainty in our TRGB magnitude
originating from the tilt of the LMC should be negligible.

\section{Discussion and Conclusion}

Using the Large Magellanic Cloud Survey by Zaritsky, Harris \& Thompson (1997),
we derive a distance modulus to the LMC of $18.59 \pm 0.09$ (random) $\pm 0.16$ 
(systematic) mag.
This value is $0.09$ mag further than the conventional value adopted by 
many groups when measuring extragalactic distances using the Cepheid variable
stars' PL relation (e.g. Mould et al. 2000).
A caveat in the TRGB method currently is, however, its dependence on the
measured distances to six Galactic globular clusters.  
If we were to correct for the proposed incompleteness effects in the calibrating 
globular cluster RGB population, then our estimate of the LMC distance would 
be raised by $3-5$\%.  
On the other hand, using the RR Lyrae calibration by
Carney et al. (1992) could shorten the distance by 9\%, even though we believe
that the Lee et al. calibration that is used in this paper agrees better
with other methods.
If the TRGB calibration is based on 
a more robust, HIPPARCOS--based RR Lyrae distance scale, such as the ones 
derived by Reid (1997) or Gratton et al. (1997), it will yield
an even larger LMC distance modulus by $\sim$0.1 mag.
Our systematic uncertainty estimate of $\pm$0.16 mag agrees reasonably conservative,
and includes uncertainties in the RR Lyrae zeropoint calibration, and
possible depth and incompleteness effect.

The MCPS data used in this paper provides an opportunity to compare the
TRGB distance to the LMC with that measured using the magnitude of the red 
clump (RC) stars.  The latter is a fairly new method; in \S1, several
estimates of the LMC distances using the red clump method were listed.
Here, we are able to compare the RC distance that is determined using
the data that are on the same photometric and extinction system.
In Figure~5, the UN-reddened magnitude distribution of RC stars in the
same regions used for the TRGB method is shown.
The red clump centroid is observed at $18.056 \pm 0.003$ mag.
Using the HIPPARCOS calibration derived by Stanek \& Garnavich (1998)
which yields a red clump magnitude of $-0.23 \pm 0.03$ mag,
we obtain a distance modulus to the LMC of $18.29 \pm 0.03$ mag.
Stanek et al. (1998) had obtained the distance modulus of 18.07 mag,
using the MCPS data and red clump method.  The difference of $\sim$0.2 mag
between this previous work and our results comes solely from the revised
reddening map (Zaritsky 1999) which is used in our analysis.
Udalski's (1998) metallicity correction (0.09[Fe/H]) would increase 
the distance modulus to 18.36.
The TRGB and RC distance estimates are inconsistent at the $\sim$5$\sigma$ level
when only internal errors are used.
A larger metallicity correction for the RC (cf. Cole 1998, Girardi et al. 1997) 
would yield better agreement.
A full analysis as to which set of corrections to apply is beyond the scope of
this paper.  However, we point out that even if the true distance modulus
to the LMC turns out to be $\sim$18.3 mag, this would not necessarily discredit
the TRGB method if the problem lies with the RR Lyrae distance scale.

We conclude by pointing out that there is excellent consistency between
the TRGB distance derived in this paper and the RR Lyrae (horizontal branch)
distances determined by several authors (e.g. Reid 1997, Gratton et al. 1997,
Carrretta et al. 1999).  This agreement is not entirely unexpected, because 
both the TRGB and RR Lyrae distances are based
on Galactic sub-dwarfs and RR Lyrae variable stars.
However, the consistency (between the TRGB and RR Lyrae distances to the LMC)
suggests that the relative magnitude offset
between the TRGB and the RR Lyrae stars matches well between
the LMC and the Galactic globular clusters which have been used to
calibrate the TRGB method (taking into account metallicity effects).  
This offers reassurance that the TRGB calibration
that has been used must be robust, implying that there cannot be large
problems, beyond the 0.1 mag level, with incompleteness effects in the 
TRGB calibration as discussed in
Cassisi \& Salaris (1997), or with the metallicity dependence.

We would like to thank Brad Gibson for providing us with an
extensive list of published distance estimates to the LMC.
S.S. acknowledges support from NASA through the  Long Term 
Space Astrophysics Program, NAS-7-1260.
DZ acknowledges financial support from an NSF grant (AST-9619576), a
NASA LTSA grant (NAG-5-3501), a David and Lucile
Packard Foundation Fellowship, and an Alfred P. Sloan
Foundation Fellowship.
RCK acknowledges the support of NSF Grants AST--9421145 and AST--9900789.

\begin{deluxetable}{lc}
\tablecolumns{2}
\tablewidth{0pc}
\tablecaption{\bf Uncertainties in the LMC Distance Modulus}
\tablehead{
\colhead{Error} &
\colhead{(mag)} \nl
}
\startdata
{\bf Random Errors} & \nl
Reddening  & 0.03 \nl
Photometry & 0.05 \nl
Tip Detection & 0.04 \nl
Color spread in the RGB population & 0.06 \nl
{\it Total} & 0.09 \nl
{\bf Systematic Errors} & \nl
RR Lyrae Distance Scale & 0.11 \nl
Undersampling in the Galactic globular cluster calibration & 0.10 \nl
Line--of--sight depth of the LMC  & 0.05 \nl
{\it Total} & 0.16 
\enddata
\end{deluxetable}

\setcounter{figure}{2}

\begin{figure}
\plotone{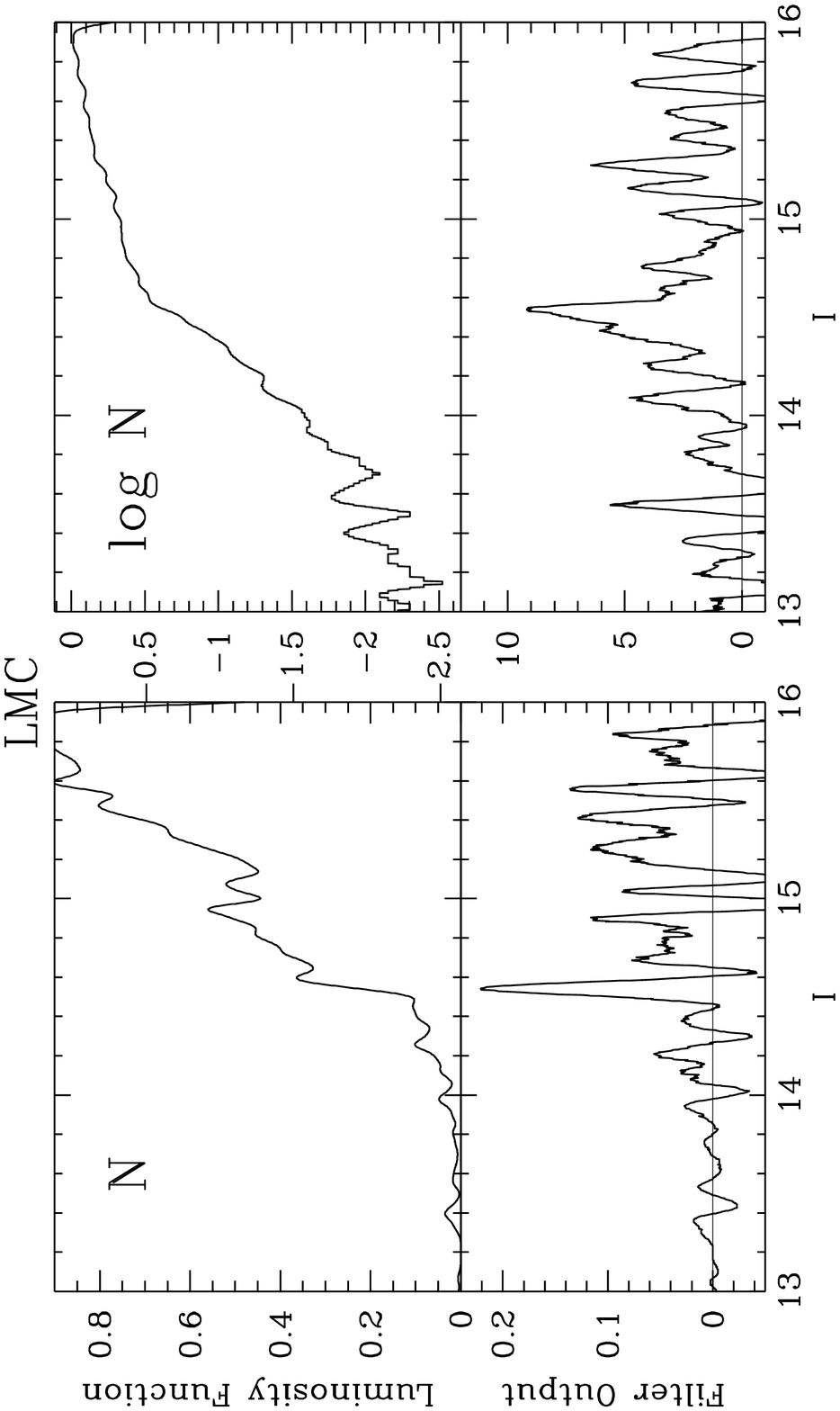}
\caption{{\it Left:} The smoothed I-band lumiosity function of the RGB
stars in the LMC (top), and its corresponding edge--detection filtering output function,
in which the location of the TRGB is indicated by the highest peak.
{\it Right:}  The logarithmic I-band luminosity function (top) and the corresponding
edge-detection filtering output.  Again, the TRGB is indicated by the highest peak,
at $M_I = 14.54$ mag.}
\end{figure}

\begin{figure}
\plotone{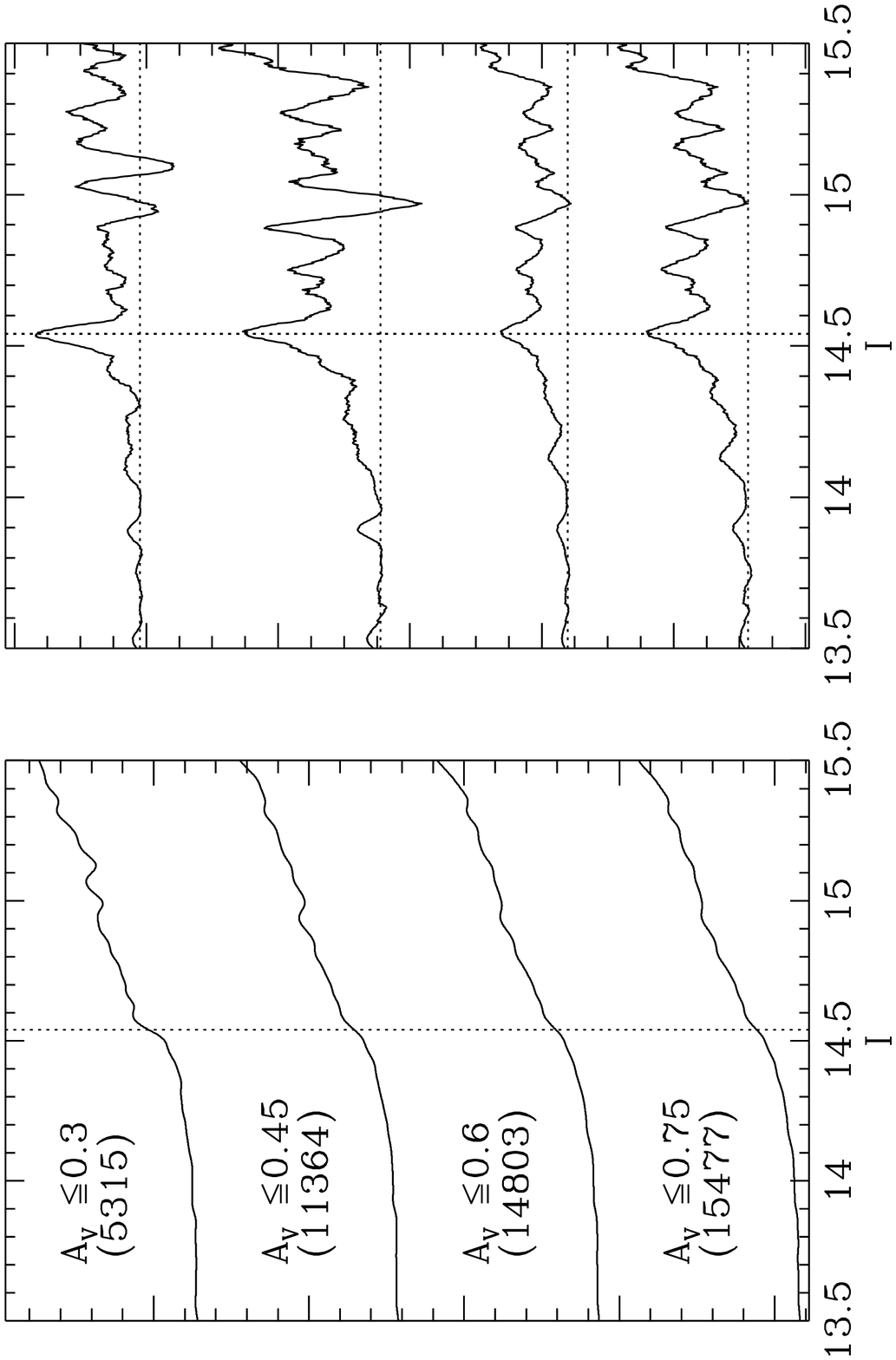}
\caption{Smoothed I-band luminosity functions (left) and filtering output
functions for four different samples for which different upper limits for the 
reddening was applied.  See text for details.
}
\end{figure}

\begin{figure}
\plotone{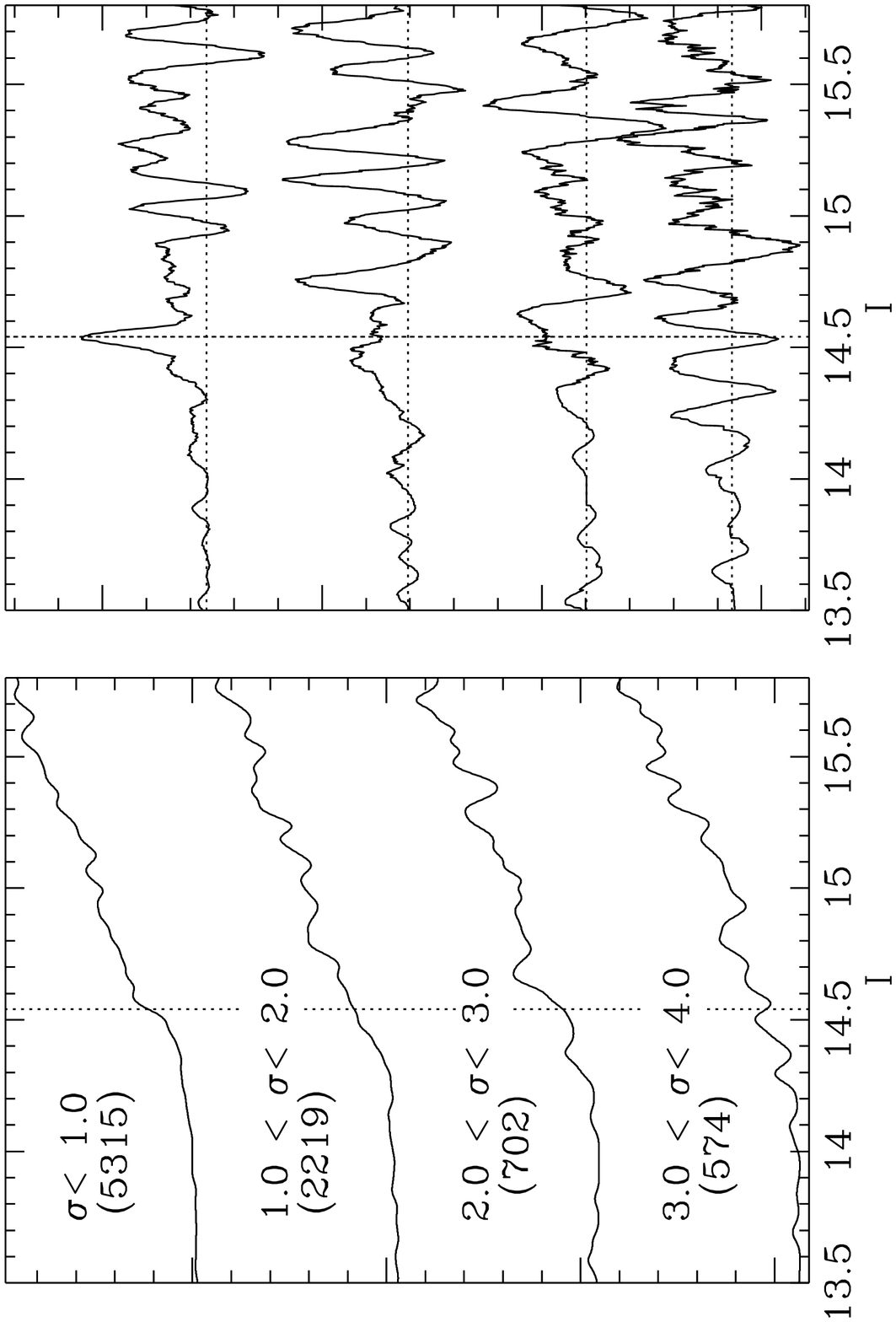}
\caption{Smoothed I-band luminosity functions (left) and filtering output
functions for four different samples for which different upper limits for the 
crowding ($\sigma$, stars/pixel) was applied.  See text for details.}
\end{figure}

\begin{figure}
\plotone{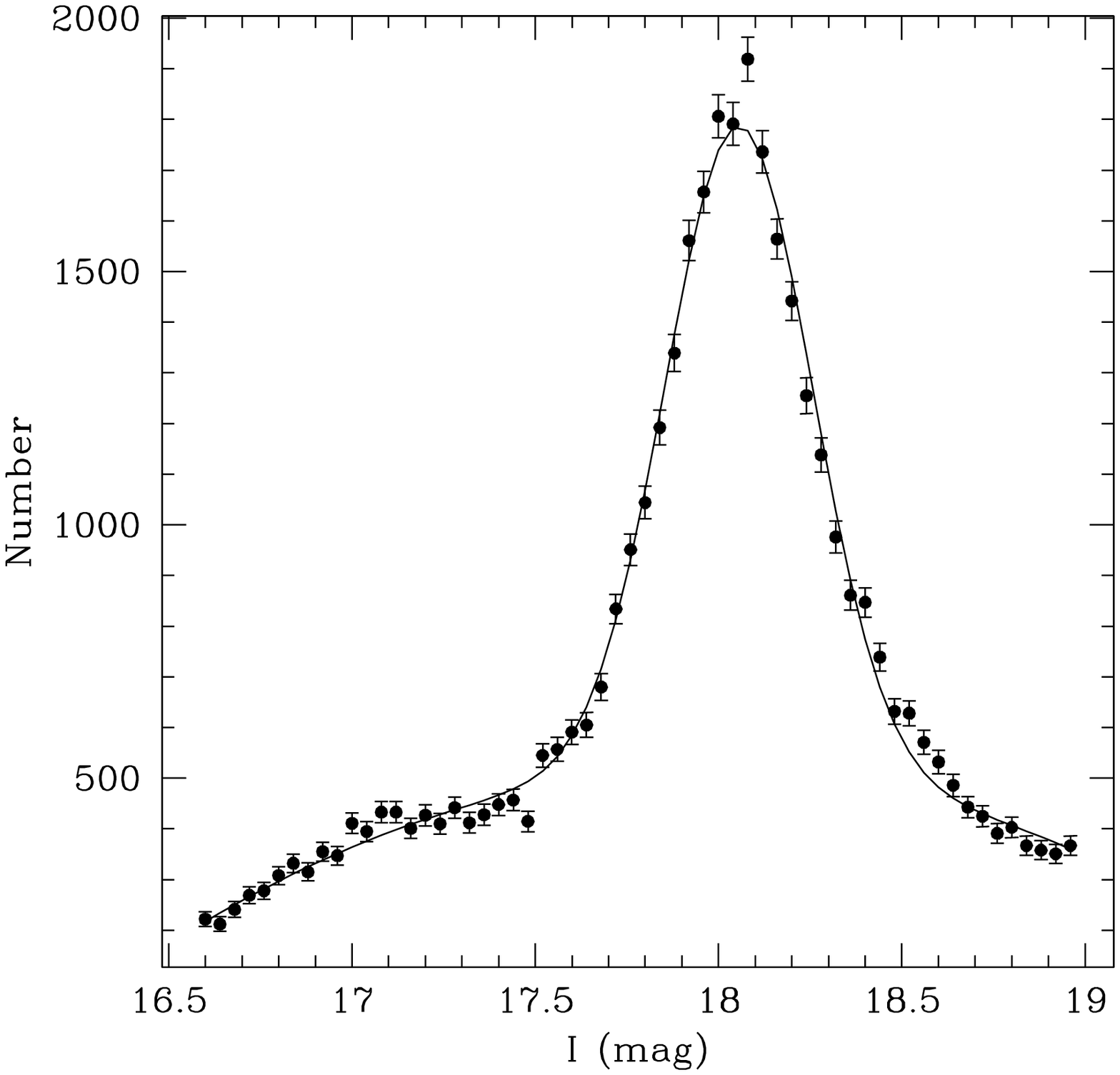}
\caption{The I-band magnitude distribution of red clump (RC) stars in the LMC.
We use stars in the same regions as used for the TRGB method.
The distribution of RC stars peaks at $18.29 \pm 0.03$ mag.
}
\end{figure}

\end{document}